\documentclass[prb,twocolumn,superscriptaddress,showpacs,floatfix]{revtex4}
\usepackage{graphicx,graphics,color,epsfig}
\usepackage{bm}
\usepackage{amsmath}
\usepackage{amssymb}
\usepackage{graphicx}

\begin{document}

\preprint{}
\title{Effects of electron-phonon interaction on non-equilibrium transport
       through single-molecule transistor}
\author{Zuo-Zi Chen}
\email{zzchen@castu.tsinghua.edu.cn} \affiliation{Center for
Advanced Study, Tsinghua University, Beijing 100084, P. R. China}
\author{Rong L\"{u}}
\affiliation{Center for Advanced Study, Tsinghua University,
Beijing 100084, P. R. China}
\author{Bang-fen Zhu}
\email{bfzhu@castu.tsinghua.edu.cn} \affiliation{Center for
Advanced Study, Tsinghua University, Beijing 100084, P. R. China}
\affiliation{Department of Physics, Tsinghua University, Beijing
100084, P. R. China}
\date{\today}

\begin{abstract}

On the basis of the nonequilibrium Green's function and
nonperturbative canonical transformation for the local
electron-phonon interaction (EPI), the quantum transport through a
single-molecule transistor(SMT) has been investigated with a
particular attention paid to the joint effect of the EPI and
SMT-lead coupling on the spectral function and conductance.  In
addition to the usual EPI-induced renormalized effects (such as
the red-shift, sharpening, and phonon-sidebands of the SMT level),
owing to improved disentagling the electron-phonon system, it has
been found that, the profile of the spectral function of the SMT
electron is sensitive to lead chemical potentials, thus can
readily be manipulated by tuning the bias as well as the SMT-gate
voltage. As a consequence, the broken particle-hole symmetry in
this system can be clearly recognized through the phonon sidebands
in the spectral function. These EPI effects also manifest
themselves in the nonequilibrium transport properties of the SMT,
particularly at low temperature.

\end{abstract}

\maketitle

\section{Introduction}

With the remarkable advances in the ability to fabricate the
nanostructures, it is possible to explore the electronic devices
based on a single molecule, i. e. the single-molecular transistor
(SMT),\cite{Park00,Reichert02} which might promise to be potential
candidates for nano-electronics and also provide with an
controllable tool in studying the fundamental physics in nanometer
scale.

It has been found that the transport through the single-molecular
transistor might exhibit profound many-body correlated effects at
low temperature, such as the Coulomb blockade or the Kondo
effect.\cite{Reichert02,Park03,Pasupathy04,Kubatkin03,Liang02} An
important feature for the SMT is its sensitivity to the local
molecular vibration, which will have profound impacts on the
transport properties.\cite{Park00,L.H. Yu04-1,L.H. Yu04-2,L.H.
Yu04-3} This may also be true in some quantum dot(QD) systems
whenever there is a strong coupling between the dot electron and
local phonon mode. In addition, the SMT is usually connected to
the outside gates via the leads. Since the applied bias voltage to
the SMT is within nanometers, a small bias might cause to
substantial field effect, thus the transport through a SMT or QD
is in general a nonequilibrium process.

Theoretically, many efforts have been made towards the quantum
transport through the SMT or QD with these features taken into
proper account. Theoretical approaches developed in this field
include the kinetic equation method,\cite{kinetic-eq} the rate
equation approach,\cite{Mitra03, Koch04} the nonequilibrium
quantum theory,\cite{Wingreen89,Gogolin02,Lundin02,Zhu03} and more
recently the numerical renormalization group
calculation.\cite{Cornaglia04-1, Cornaglia04-2,Paaske04} In doing
such investigations, several groups have already accounted for
 the electron-phonon interaction (EPI), many-body effects, or nonequilibrium properties.
\cite{correlation,Flensberg02,Koch04}
 However, different results may be obtained by
different treatments of these effects.\cite {Koch04,Zhu03}
Although the numerical renormalization group method can well
predict the equilibrium properties of the system, it cannot be
applied to the nonequilibrium situation directly. While regarding
to the present systems, a theory capable of dealing with both
nonequilibrium and strong EPI situations should be favorite. One
candidate, the Keldysh Green's function, has shown to be
successful for the nonequilibrium systems.\cite{Huag96,
Langreth76, David02,Caroli72,Anda91,Galperin04} Within this
framework the nonequilibrium Dyson equations are solved
self-consistently, in which the self energies due to the EPI can
be calculated by the standard many-body diagrammatic
technique,\cite{Caroli72,Anda91} or by using the self-consistent
Born approximation.\cite{Galperin04} Such an approach can meet the
present concerns, and can be generalized to account for the
effects arising from the phonon dynamics itself.\cite{Galperin04}
But, in general, the calculations are complicate as
self-consistent integrals are encountered. This difficulty can be
circumvented by directly diagonalizing the EPI terms with a
canonical transformation,\cite{Mahan00,Hewson80} as long as the
phonons are in the thermodynamic equilibrium and the fluctuation
in phonons can be neglected. By an improved treatment in the
decoupling approximation as presented in the present work, which
differs from the majority of previous publications,\cite{Zhu03,
Mahan00} the electron can be properly decoupled from phonon
system, and yet permits to straightforward solutions in a compact,
analytic form.

In this paper, by combining the nonequilibrium Keldysh Green's
function technique with the canonical transformation for the
electron-phonon system, we intend to study  the quantum transport
through the SMT in the presence of finite bias and strong local
EPI, with special attention paid to the effect of the improved
decoupling treatment for the electron-phonon system on the
spectral function. The paper is organized as follows. In section
II, an Anderson-Holstein model is introduced for the SMT in the
presence of local EPI, then our main theoretical framework is
described, in which the commonly used procedure for decoupling the
electron-phonon system is carefully reexamined. In section III,
the numerical results for the spectral function of the SMT
electron is demonstrated and discussed, including both the
dressing effects due to the EPI and the joint effects caused by
its interplay with the SMT-leads coupling. An interesting
phenomenon is predicted that the profile of the spectral function
for the SMT electron can be manipulated at low temperature by
tuning the chemical potentials in the leads. Consequently, the
phonon sidebands in the spectral function may be quite asymmetric
with respect to the renormalized SMT level, indicating a broken
electron-hole symmetry. In section IV, it is shown how the effect
of the different decoupling approximation on the spectral function
manifests itself in the conductances of the SMT at zero
temperature and finite temperatures, which might yield nearly the
same results at high temperature, but definitely different
behaviors at zero temperature. Finally, a brief conclusion is
drawn.

\section{Physical Model and Formalism}

\begin{figure}[tbp]

\includegraphics[width=8.5cm]{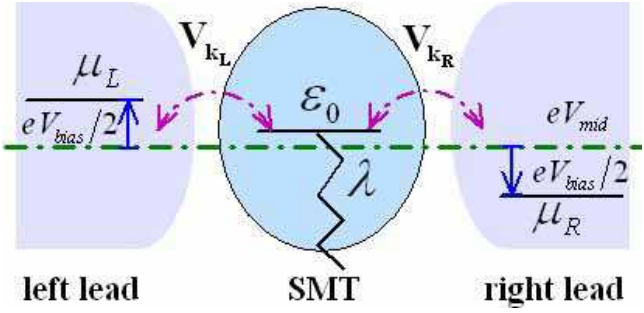}
\caption{Schematic illustration for the single-molecule transistor
system. } \label{configure}
\end{figure}

\subsection{Model}
As shown in Fig.\ref{configure}, our model system is a single
electron level in a SMT or QD, which is coupled to the local
vibration mode as well as to two metallic leads.  For the sake of
simplicity, we shall restrict ourselves exclusively to the effect
of the EPI and chemical potentials in two leads, thus ignore other
factors, such as the intricacies of the real SMT, Coulomb
interaction and spin effect. The model Hamiltonian can be then
expressed as
\begin{equation}
{\bf H}={\bf H}_{\text{leads}}+{\bf H}_{\text{ph}}+{\bf
H}_{\text{D}}+{\bf H}_{\text{T}},
\end{equation}
where the first two terms of the Hamiltonian represent the
noninteracting electron gas in the leads and the local vibration
mode of the SMT, respectively, namely,
\begin{equation}
{\bf H}_{\text{leads}}=\sum_{{\bf k}\in L(R)}\epsilon _{\bf k}{\bf
c}_{\bf k }^{\dagger }{\bf c}_{\bf k},\qquad{\bf
H}_{\text{ph}}=\omega _{0}{\bf a}^{\dag }{\bf a}.
\end{equation}
Here the operator ${\bf c}_{\bf k }^{\dagger }$ $\left( {\bf
c}_{\bf k }\right) $ creates (annihilates) a conduction electron
with momentum ${\bf k}$ and energy $\epsilon _{\bf k}$ in the left
(L) or the right (R) lead, $\omega _{0}$ is the vibrational
frequency of the molecule, and ${\bf a}^{\dag } $ $\left( {\bf
a}\right) $ is the phonon creation (annihilation) operator. The
third term ${\bf H}_{D}$ describes the coupling between the SMT
electron and local phonon mode with strength $\lambda$,
\begin{equation}
{\bf H}_{\text{D}}=\left[ \epsilon _{0}+\lambda \left( {\bf a}+{\bf a}^{\dag }\right) %
\right] {\bf d}^{\dag }{\bf d},\label{H_D}
\end{equation}
where  ${\bf d}^{\dag }$ $\left( {\bf d}\right) $ is the
corresponding creation (annihilation) operator of the SMT or QD
electron associated with the energy $\epsilon_0$. The last term
${\bf H}_{T}$ describes the hopping of electron between the SMT
and leads with the tunnelling matrix elements denoted as $V_{\bf
k}$,
\begin{equation}
{\bf H}_{T}=\sum_{{\bf k}\in L(R)}\left[ V_{\bf k}{\bf c}_{\bf k
}^{\dagger }{\bf d}+ \text{h.c.}\right].
\end{equation}

The chemical potentials in the left and right lead are denoted by
$\mu_{L(R)}$, respectively, which are related to the bias,
$V_{bias}$, and the average of two lead potentials, $V_{mid}$,
through $V_{bias}=(\mu_L-\mu_R)/e$ and
$V_{mid}\equiv(\mu_L+\mu_R)/2e$. By changing the external bias and
gate voltages experimentally, $\mu_{L(R)}$ and $\epsilon_0$ can be
adjusted independently.

In the strong EPI regime, it is appropriate to eliminate the
electron-phonon coupling terms in the Hamiltonian by using a
nonperturbative canonical transformation, {\it i.e.},
$\overline{\bf H}=e^{\mathcal{\bf S}}{\bf H}e^{-\mathcal{\bf S}}$
with $\mathcal{\bf S}=\frac{\lambda }{\omega _{0}}{\bf
 d}^{\dag }{\bf d}\left( {\bf a}^{\dag }-{\bf
 a}\right)$.
The transformed Hamiltonian reads
$\overline{\bf H}=\overline{\bf H}_{\text{ph}}+\overline{\bf
H}_{\text{el}}$, where the phonon part keeps unchanged, while the
electron part is reshaped into
\begin{eqnarray} \overline{\bf
H}_{el}&=&\sum_{\bf k}\epsilon _{\bf k}{\bf c}_{\bf k}^{\dagger
}{\bf c}_{\bf k }+\widetilde{ \epsilon }_{0} {\bf d}^{\dag }{\bf
d}+\sum_{\bf k}\left[ \widetilde{V}_{\bf k}{\bf c}_{\bf k
}^{\dagger }{\bf d}+\text{h.c.}\right].\label{H_decoupled}
\end{eqnarray}

It is clear that due to the EPI, the energy level of SMT is
renormalized to $\widetilde{ \epsilon }_{0}\equiv\epsilon
_{0}-g\omega_0$, where $ g\equiv\left( \lambda /\omega _{0}\right)
^{2}$, and the dressed tunnelling matrix elements are transformed
into $\widetilde{V}_{\bf k }\equiv{V}_{\bf k }{\bf X}$, where the
phonon operator ${\bf X}\equiv\exp \left[ -\left( \lambda /\omega
_{0}\right) \left( {\bf a}^{\dag }-{\bf a}\right) \right] $ arises
from the canonical transformation of the particle operator
$e^{\mathcal{\bf S}}{\bf d}e^{-\mathcal{\bf S}}={\bf d X}$. This
reveals that the interaction between the SMT electron and the
local phonon mode results in an effective phonon-mediated coupling
between the SMT and the lead electrons. As in dealing with the
localized polaron \cite {Hewson80, Mahan00}, here it is reasonable
to replace the operator ${\bf X}$ with its expectation value
$\langle {\bf X} \rangle$, i.e., $\widetilde{V}_{\bf k }=V_{\bf k
}\exp \left[ -g\left( N_{ph}+1/2\right) \right] $, where $N_{ph}$
is the phonon population, and can be expressed as $N_{ph}=1/\left[
\exp \left(\beta \omega _{0}\right) -1\right]$ with $\beta
=1/k_{B}T$. Notice that this is an important approximation made in
the present paper, which is valid only when the hopping is small
compared to the EPI, i.e. $V_{\bf k}\ll\lambda$.

\subsection{Formalism}
By using the Keldysh Green's function technique\cite{Huag96,
Mahan00, Langreth76}, the current through an interaction region
coupled to two leads can be expressed as\cite{Meir94,Meir92},
\begin{eqnarray}
J =\frac{ie}{2h}\int d\omega \left\{\left[ f_{L}\left( \omega
\right)\Gamma^{L}-f_{R}\left( \omega \right) \Gamma
^{R}\right] \left( G^{r}(\omega)-G^{a}(\omega)\right)  \right.  \notag \\
\left.+\left( \Gamma^{L}-\Gamma^{R}\right) G^{<}(\omega)
\right\},\label{J}
\end{eqnarray}%
where $f_{L\left( R\right) }$ is the Fermi distribution function
in the left (right) lead, $\Gamma^{L\left( R\right)
}(\epsilon)\equiv2\pi \rho_{L(R)}(\epsilon)\left\vert
V_{L(R)}(\epsilon)\right\vert ^2$, $\rho_{L(R)}(\epsilon)$ is the
density of states in the left (right) lead, $V_{L(R)}(\epsilon)$
equals $V_{{\bf k}\in L(R)}$ for $\epsilon=\epsilon_{\bf k}$, and
$G^{r(a)}(\omega)$, $G^{<}(\omega)$ are the Fourier
transformations of the standard Keldysh retarded (advanced) and
lesser Green functions for the dot electron,
respectively\cite{Mahan00}. To note that the parameters
$\Gamma_{L(R)}$ are not the renormalized ones, because derivation
of this current formula only relies on the hopping terms,
regardless the nature of the interacting region. In calculating
the current, one needs the spectral function, which is defined as
\begin{eqnarray}
A(\omega)= i\left( G^{r}(\omega)-G^{a}(\omega)\right) =i\left(
G^{>}(\omega)-G^{<}(\omega)\right),
\end{eqnarray}
and the lesser Green's function $G^<$, which is proportional to
spectral function and the  occupation of the electron.

When the operator ${\bf X}$ is replaced by its expectation value,
the Hamiltonian Eq.(\ref{H_decoupled}) is decoupled from the
phonon operator, then the interacting lesser Green function may be
separated:
\begin{eqnarray}
&&G^{<}\left( t\right) \equiv i\left\langle {\bf d}^{\dag }\left(
0\right) {\bf d}\left( t\right) \right\rangle=i\left\langle
\overline{\bf d}^{\dag }e^{i\overline{\bf H}t} \overline{\bf
d}e^{-i\overline{\bf H}t}
\right\rangle\nonumber\\
&=& i\left\langle {\bf d}^{\dag }e^{i\overline{\bf H}_{el}t} {\bf
d}e^{-i\overline{\bf H}_{el}t} \right\rangle_{el}\left\langle {\bf
X}^{\dag}e^{i\overline{\bf H}_{ph}t}{\bf X}e^{-i\overline{\bf
H}_{ph}t}
\right\rangle_{ph}\nonumber\\
&\equiv&\widetilde{G}^{<}\left( t\right) e^{-\Phi \left( -t\right)
},\label{approximation_G^>}
\end{eqnarray}
and similarly,
\begin{eqnarray}
G^{>}\left( t\right)&\equiv& -i\left\langle {\bf d}\left(
t\right){\bf d}^{\dag }\left( 0\right)
\right\rangle=\widetilde{G}^{>}\left( t\right) e^{-\Phi \left(
t\right) }\label{approximation_G^<},
\end{eqnarray}
where $\widetilde{G}^{>(<)}\left( t\right)$ is the dressed greater
(lesser) Green function for a dressed SMT electron governed by
$\overline{\bf H}_{el}$, and the factor $e^{-\Phi \left( \mp
t\right) }$ arises from the trace of the phonon parts
$\left\langle {\bf X}^{\dag}(0){\bf X}(t) \right\rangle_{ph}$ or
$\left\langle {\bf X}(t){\bf X}^{\dag}(0) \right\rangle_{ph}$,
respectively, \cite{Mahan00}
\begin{eqnarray}
\Phi \left( t\right)=g \left[ N_{ph}\left( 1-e^{i\omega
_{0}t}\right) +\left( N_{ph}+1\right) \left( 1-e^{-i\omega
_{0}t}\right) \right].\label{phi}
\end{eqnarray}
Note that because $\Phi(-t)\neq\Phi(t)$, the decoupling
approximation used in the majority of previous
publications,\cite{Zhu03, Huag96} which directly decouples the
retarded (advanced) Green function as
\begin{eqnarray}
G^{r(a)}(t)=\widetilde{G}^{r(a)}(t)e^{-\Phi \left( t\right)
},\label{approximation_G^r}
\end{eqnarray}
has ignored the difference between the $N_{ph}$ and $N_{ph}+1$ in
the expression of $\Phi(t)$. Although such approximation works
well at high temperature, because $N_{ph}\approx N_{ph}+1$, it
does make difference at low temperature because of vanishing
$N_{ph}$. We shall further discuss this point latter.

By using the identity $e^{-\Phi \left( t\right) }=\sum_{n=-\infty
}^{\infty }L_n e^{-in\omega _{0}t}$, the greater and lesser Green
functions can be respectively expanded as
\begin{eqnarray}
G^>(\omega)&=&\sum_{n=-\infty}^{\infty}L_{n}\widetilde{G}^{>}\left(
\omega -n\omega _{0}\right),\notag\\
G^<(\omega)&=&\sum_{n=-\infty}^{\infty}L_{n}\widetilde{G}^{<}\left(
\omega +n\omega _{0}\right),\label{G^<>}
\end{eqnarray}
where the index $n$ stands for the number of phonons involved, and
$L_n$ are the coefficients depending on temperature and the
strength of EPI. At finite temperature,
\begin{eqnarray}
L_n\equiv e^{-g\left( 2N_{ph}+1\right)
}e^{n\omega_0\beta/2}I_{n}\left( 2g\sqrt{N_{ph}\left(
N_{ph}+1\right) }\right),\label{Ln}
\end{eqnarray}
where $I_n(z)$ is the $n$-th Bessel function of complex argument. At
zero temperature, $L_n$ simply reads
\begin{eqnarray}
L_n\equiv\left\{\begin{tabular}{cc}
 $e^{-g}\frac{g^n}{n!}$&$n\geq0$\\
 $0$&$n<0$\\
\end{tabular}\right.\label{Ln0}
\end{eqnarray}
Thus the spectral function can be expressed as
\begin{eqnarray}
A(\omega)&=&\sum_{n=-\infty }^{\infty
}iL_{n}\left[\widetilde{G}^>(\omega -n\omega
_{0})-\widetilde{G}^<(\omega+n\omega_0)\right]. \label{A}
\end{eqnarray}

\begin{figure}[tbp]
\includegraphics[width=8.5cm]{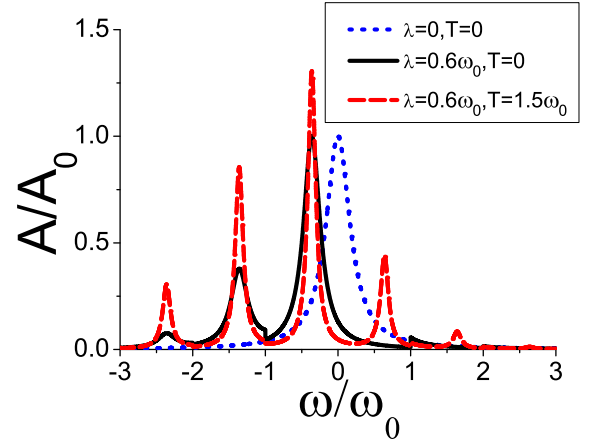}
\caption{The dimensionless Spectral function of the SMT electron
for different EPI strengths and temperatures. The parameters for
calculation are taken as: $\Gamma=0.2\omega_0$,
$\mu_L=\mu_R=\epsilon_0=0$, and the unit $A_0=2/\Gamma$. The
strength of the EPI used here is the same as the references
\cite{David02}.} \label{A-lambda-T}
\end{figure}

With the help of the equation of motion approach, the retarded
(advanced) Green function for the dressed electron can be
analytically evaluated as
\begin{equation}
\widetilde{G}^{r\left( a\right) }\left( \omega \right) =\frac{1}{%
\omega -\widetilde{\epsilon
}_{0}-\widetilde{\Sigma}^{r(a)}(\omega)},\label{G^r_tilde}
\end{equation}
where the retarded (advanced) self-energy is
\begin{eqnarray}
\widetilde\Sigma^{r(a)}(\omega)=\sum_{k\in
L,R}\frac{|\widetilde{V}_{k}|^{2}}{\omega -\epsilon _{k}\pm
i\eta}=\widetilde{\Lambda}(\omega)\mp i\widetilde{\Gamma}(\omega).
\end{eqnarray}
For simplicity, in the wide-band limit, both the real and the
imaginary part of the self-energy, $\widetilde{\Lambda}(\omega)$,
and $\widetilde{\Gamma}(\omega)$, are assumed to be constants
independent of energy. Thus, when the real part of the self-energy
is absorbed into the SMT level shift, only the broadening due to
the tunnel coupling will be considered. With the assumption of
symmetric coupling,
$\widetilde{\Gamma}^L\approx\widetilde{\Gamma}^R=\widetilde{\Gamma}$,
the broadening can be expressed as
$\widetilde{\Gamma}(\omega)=(\widetilde{\Gamma}^L+\widetilde{\Gamma}^R)/2=\widetilde{\Gamma}$.
Then the spectral function of the dressed SMT electron can be cast
into
\begin{eqnarray}
\widetilde{A}(\omega)= \frac{2\widetilde{\Gamma }}{\left( \omega
-\widetilde{\epsilon }_{0}\right) ^{2}+ \widetilde{\Gamma }
^{2}}\label{A_tilde}.
\end{eqnarray}
Following the Keldysh equation \cite {Langreth76} for the lesser
Green function, {\it i.e.},
$\widetilde{G}^{<}(\omega)=\widetilde{G}^{r}(\omega)\widetilde{\Sigma}^{<}(\omega)\widetilde{G}^{a}(\omega)$
with $\widetilde{\Sigma}^{<}(\omega)=
i\widetilde{\Gamma}(f_L(\omega)+f_R(\omega))$, the relation
between the dressed greater or lesser Green function and the
dressed spectral function is respectively as
\begin{eqnarray}
\widetilde{G}^{>}\left( \omega \right) &=&-i\frac{2- f_{L}\left(
\omega \right)-f_{R}\left( \omega
\right)}{2}\widetilde{A}(\omega),\notag\\
\widetilde{G}^{<}\left( \omega \right)&=&i\frac{f_{L}\left( \omega
\right)+f_{R}\left( \omega \right)}{2}
\widetilde{A}(\omega)\label{G^<>_tilde}.
\end{eqnarray}
Thus the spectral function of the SMT electron $A(\omega)$ can
readily be obtained by substituting Eqs.(\ref{G^<>_tilde}) and
(\ref{A_tilde}) into Eq.(\ref{A}).

\section{The spectral function of the SMT electron}

\begin{figure}[tbp]
\includegraphics[width=8.5cm]{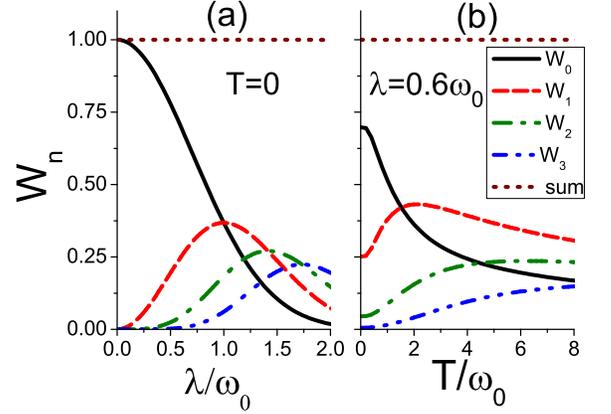}
\caption{The total weights of the $+n$-th and the $-n$-th phonon
sidebands, $W_n$, as functions of the (a) strength of EPI
$\lambda$ and (b) temperature $T$, where $\Gamma=0.2\omega_0$,
while $\mu_{L(R)}$ are arbitrary. The dot lines are the numerical
summation, $\sum_n W_n$.} \label{Wn}
\end{figure}

\subsection{The dressing effects caused by the EPI}
The spectral function for our structure generally depends on the
 EPI strength, temperature, and chemical potential in two leads
relative to the SMT level. Let us first address the first two
dependencies in the case of $\mu_L=\mu_R=\epsilon_0$.

As plotted in Fig.\ref{A-lambda-T}, the effect of the EPI and
temperature on the spectral functions of the SMT electron are
evaluated. For comparison, the spectral function without the EPI
at $0K$ is also shown, which exhibits a single resonant peak at
$\epsilon_0$ with a Lorentzian. Compared with the EPI-free case,
with finite EPI, the resonant peak at $\epsilon_0$ in the spectral
function is red-shifted by $g\omega_0$, and the spectral peak is
sharpened. This results from the renormalized effects on the SMT
level and the dressing effect on the tunnelling matrix elements
$V_{\bf k}$ due to the EPI. More noticeably, new satellite peaks
may appear at $\widetilde{\epsilon}_0\pm n\omega_0$ in the
spectral function, implying that due to the EPI the ground state
for the coupled system may possess finite components involving $n$
phonons. For later convenience, we label the resonant peak located
at $\widetilde{\epsilon}_0$ as the zero-phonon-band, and the
satellite peaks located at $\widetilde{\epsilon}_0\pm n\omega_0$
as the $\pm n$-th phonon sidebands. In general, these phonon
sidebands are not symmetric with respect to
$\widetilde{\epsilon}_0$. By using the identity
\begin{eqnarray}
i\int_{-\infty}^{\infty}d\omega(\widetilde{G}^>(\omega-n\omega_0)-\widetilde{G}^<(\omega+n\omega_0))\notag\\
=i\int_{-\infty}^{\infty}d\omega(\widetilde{G}^>(\omega)-\widetilde{G}^<(\omega))
=2\pi,
\end{eqnarray}
and $\sum_{n=-\infty}^{\infty}L_n=\exp\{\Phi(0)\}=1$, it can be
seen that the sum rule for the spectral function still
holds in the presence of EPI, namely
\begin{eqnarray}
\int_{-\infty}^{\infty}{d\omega}{A}(\omega)={2\pi}.\label{sum_A}
\end{eqnarray}

As shown in Fig.\ref{A-lambda-T}, the weight of each peak,
defined as the integrated area under the peak divided by $2\pi$
and denoted by $W_{\pm n}$, is sensitive to the EPI strength and
temperature. We can further define sum of the spectral
weights of the $+n$-th and $-n$-th phonon sidebands as
\begin{eqnarray}
&&W_n\equiv W_{+n}+W_{-n}=L_n+L_{-n}=e^{-g\left( 2N_{ph}+1\right)
}\notag\\
&\times&(e^{n\omega_0\beta/2}+e^{-n\omega_0\beta/2})I_{n}\left(
2g\sqrt{N_{ph}\left( N_{ph}+1\right) }\right).\label{W_n}
\end{eqnarray}
Fig.\ref{Wn} shows that when $\lambda$ or $T$ increases, the
spectral weight of the zero-phonon band, $W_0$ decreases
monotonically; while the phonon sideband, $W_n(n\neq0)$, gets
lager first, then reaches a maximum value for a certain EPI
strength or temperature, and decreases again. This reflects that
when increasing the temperature or EPI strength, the occupation
probability of the phonon sideband increases, keeping the
conservation of the total spectral weight, $\it i.e.$,
$\sum_{n=0}^{\infty}W_n=1$. Note that these variation trends are
not in contradiction with Fig.\ref{A-lambda-T}, where the heights
of both the zero-phonon peak and phonon sidebands increase as
temperature gets higher, as the corresponding widths are narrowed
simultaneously.

\subsection{The interplay between the EPI and the SMT electron-lead coupling}
Now let us take a close look at the dependence of the spectral
function of the SMT electron on the chemical potentials in two
leads in the presence of the EPI.

\begin{figure}[tbp]
\includegraphics[width=4.0cm]{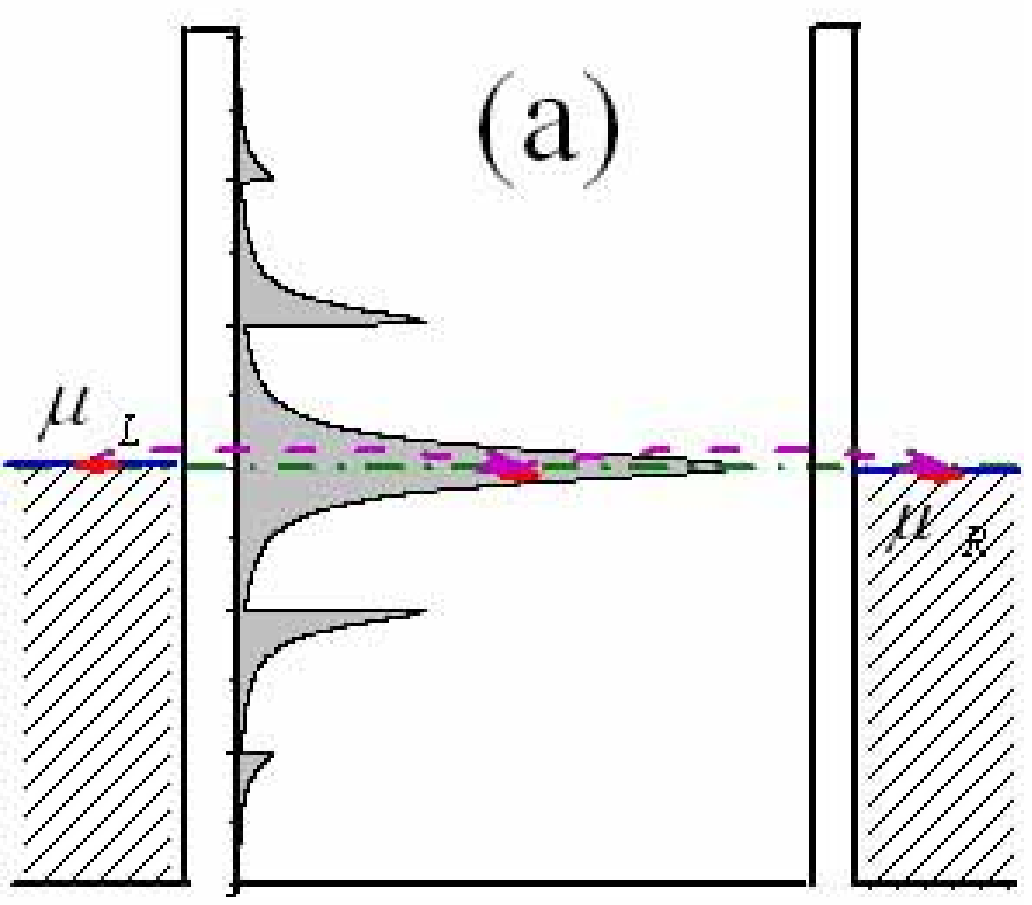}
\includegraphics[width=4.0cm]{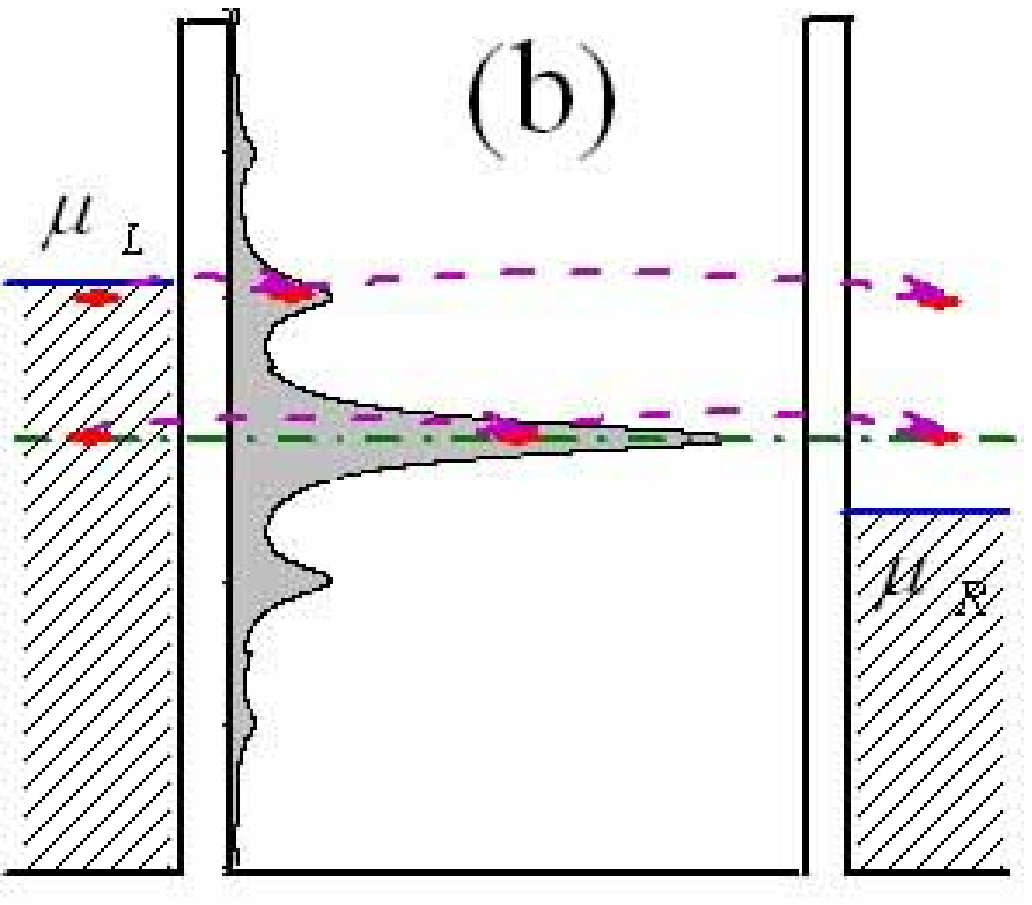}
\includegraphics[width=4.0cm]{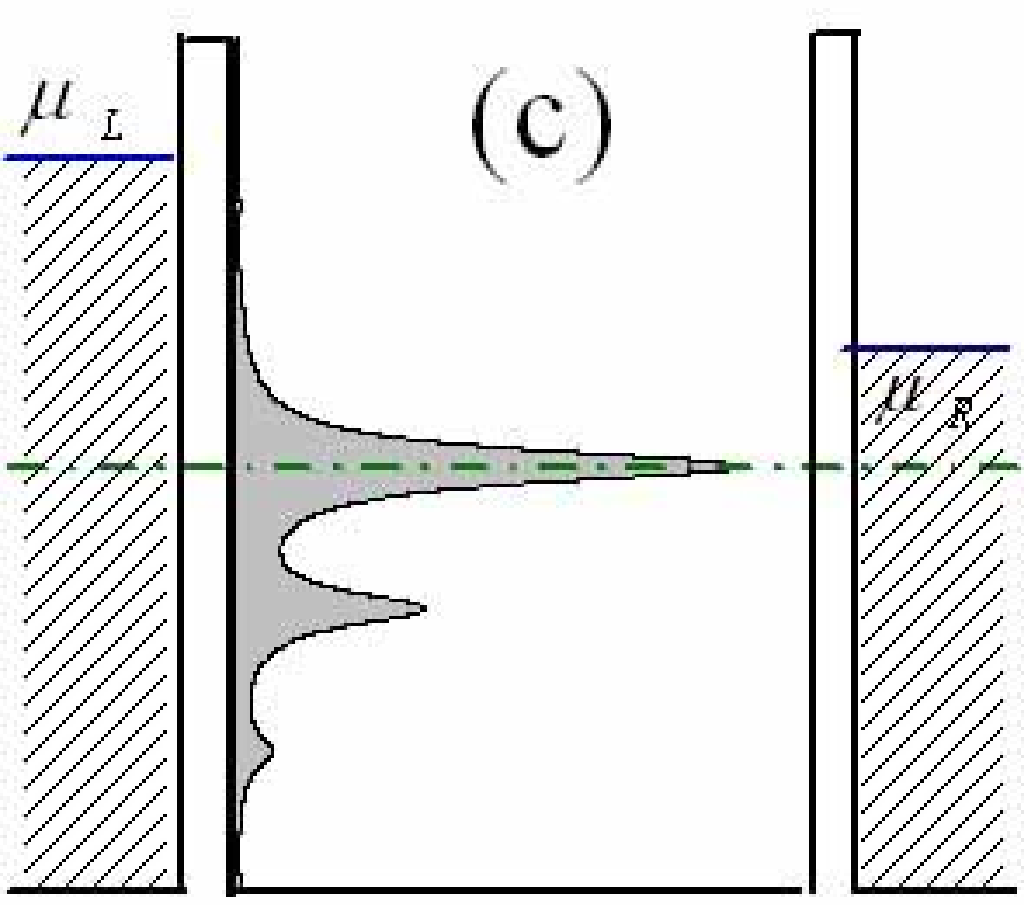}
\includegraphics[width=4.0cm]{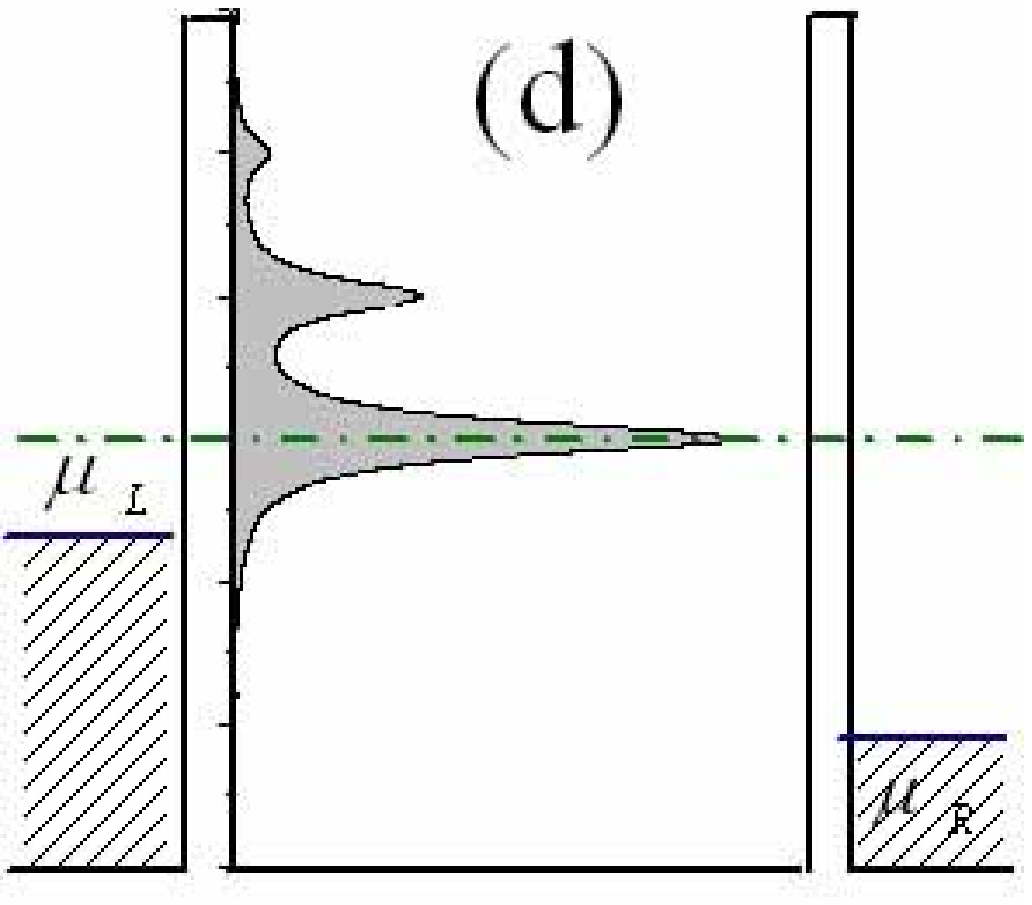}
\caption{The spectral functions of the SMT electron for four
typical configurations of $\mu_{L(R)}$: (a)
$\mu_L=\mu_R=\widetilde{\epsilon}_0$, (b)
$\mu_L=\widetilde{\epsilon}_0+0.9\omega_0,
\mu_R=\widetilde{\epsilon}_0-0.5\omega_0$, (c)
$\mu_L=\widetilde{\epsilon}_0+2.2\omega_0,
\mu_R=\widetilde{\epsilon}_0+0.8\omega_0$, (d)
$\mu_L=\widetilde{\epsilon}_0-0.8\omega_0,
\mu_R=\widetilde{\epsilon}_0-2.2\omega_0$, where the parameters
are $\Gamma=0.2\omega_0$, $\lambda=0.6\omega_0$, $T=0$ and
$\widetilde{\epsilon}_0=0$. The balls connected by the dash lines
in (a) and (b) denote the resonant tunnelling processes for
electrons.} \label{A-mu}
\end{figure}

As shown in Fig.\ref{A-mu}, it is quite interesting that the
spectral function of the SMT electron profiles quite differently
for the same EPI strength and temperature ($0K$), but under
different configurations of the chemical potentials in two leads.
And the phonon sidebands, whose lineshapes may vary abruptly at
certain frequency, can be nonvanishing in both sides of the
zero-phonon band, even at zero temperature. This is quite
different from the usual Independent Boson Model,
according to which, the phonon sidebands have nothing to do with
the lead chemical potentials, and at zero temperature, they should
vanish on the negative energy side of the zero-phonon band of the hole.

The unusual behavior of the present spectral function can be
understood with Eqs.(\ref{A}) and (\ref{G^<>_tilde}), in which
 $\widetilde{G}^>(\omega-n\omega_0)$, and
$\widetilde{G}^<(\omega+n\omega_0)$ in general depends on the
Fermi function of $f_L(f_R)$ in the left (right) lead, and thus
on the chemical potential, $\mu_L(\mu_R)$, respectively. At zero
temperature, the expansion coefficients $L_n$ vanish for negative $n$, therefore
only the zero-phonon band of the spectral function can both the
lesser and greater Green function contribute to; while the
phonon sideband below the zero-phonon band results from the
$\widetilde{G}^<(\omega+n\omega_0)$, and that above the
zero-phonon band comes from the
$\widetilde{G}^>(\omega-n\omega_0)$.

Recalling that the lesser and greater Green functions
$\widetilde{G}^{<(>)}$ correspond to the dressed SMT electron and
hole propagator, which are proportional to  the occupation number,
$n_{SMT}$, for the SMT electron, or $1-n_{SMT}$ for the hole,
respectively. Different from the Independent Boson Model, where
$n_{SMT}$ is fixed to either zero or one at $0K$, as the Fermi
level for the isolate system is well defined, in the present
non-equilibrium model, due to the tunnelling coupling the
occupation of the SMT electron is determined by the chemical
potentials at the left and right leads when the system approaches
a steady state. Thus, by tuning the lead chemical potentials,
$n_{SMT}$ can take any value between zero and one, consequently,
even at $0K$ the contribution from both the electron and hole to
the spectral function of the SMT need considering. At low
temperature the SMT electron and hole can only emit phonons, so
the phonon sidebands below the zero-phonon band come from the
occupative SMT electron, while those above the zero-phonon band
originate from the SMT hole. If there is a partial occupation in
the zero-phonon band for the SMT electron, it is required to
carefully examine the contribution from both electron and hole. It
is the different treatment of $n_{SMT}$ that leads to the unusual
behavior of the phonon sidebands at zero temperature.

The spectral function can be explicitly expressed as functions of
the bias and $V_{mid}$, the average of two lead potentials relative to the
renormalized SMT level as defined in the Section II,
 {\it i.e.}, $A(\omega,V_{mid},V_{bias})$. In this way the effect
due to the bias and due to the $V_{mid}$ on the spectral function
will be considered separately. We have found that the $V_{mid}$
mostly affects the lineshape of the spectral function in the
present model. Four typical configurations of the lead chemical
potentials in Fig.\ref{A-mu} are divided into two categories: (1)
The $V_{mid} \sim \widetilde{\epsilon}_0$, namely the SMT level is
partially filled. Then for both bias
 in Fig.\ref{A-mu}(a) and Fig.\ref{A-mu}(b), the resonant
tunnelling may take place. (2) The $V_{mid}$ deviates from
$\widetilde{\epsilon}_0$ significantly compared with the
band-width $\widetilde{\Gamma}$, so that the SMT level is either
fully occupied or totally empty, {\it i.e.} $n_{SMT}\simeq1$
(Fig.\ref{A-mu}(c)), or $n_{SMT}\simeq0$ (Fig.\ref{A-mu}(d)), in
which no resonant tunnelling can occur.

In Fig.\ref{A-mu}(a), both chemical potentials in the leads align
exactly with the renormalized SMT level, where and the lineshape
of each phonon sideband exhibits discontinuity at certain
frequency. This is similar to the case $\mu_L=\mu_R=\epsilon_0$
({\it cf.} Fig.\ref{A-lambda-T}), except for that the spectral
function in Fig.\ref{A-mu}(a) is symmetric with respect to
$\widetilde{\epsilon}_0$, while that in Fig.\ref{A-lambda-T} is
asymmetric. This is understandable, because at zero bias and zero
temperature there exist a well-defined boundary, $\mu_L=\mu_R$,
separating the SMT electron from hole. The symmetric spectral
function is an exception when the renormalized SMT level happens
to coincide with the boundary and $n_{SMT}=0.5$. In
Fig.\ref{A-mu}(b), when the bias is large enough to enclose the
most part of the zero-phonon band, both distributions of the SMT
electron or hole in the zero-phonon band become Lorenzian, and so
does the lineshape of each phonon sideband. Strictly speaking, the
spectral function in Fig.\ref{A-mu}(b) is not symmetric around
$\widetilde{\epsilon}_0$ as $V_{mid}\neq\widetilde{\epsilon}_0$,
but the asymmetric part can be hardly discerned for the present
case. It is noticed that whenever one of the phonon sidebands
enters the bias region, there will be the phonon-assisted
tunnelling process. Compare Fig.\ref{A-mu}(b) to (a), it is
evident that the bias is not only important in determining the
profile of each phonon sideband, but also crucial in controlling
the phonon-assisted tunnelling. It is interesting to observe that
when $n_{SMT}\simeq0$ (Fig.\ref{A-mu}(d)), which corresponds to
the one-hole picture, the usual result following the Independent
Boson Model with one hole\cite{Mahan00} is recovered; on the other
hand, when $n_{SMT}\simeq1$ as in Fig.\ref{A-mu}(c), which
corresponds to the picture of one-electron, the spectral function
is reversed with respect to $\widetilde{\epsilon}_0$ compared with
(d). Comparison among the four configurations shows that the
$V_{mid}$ relative to $\widetilde{\epsilon}_0$ has played a
decisive role in determining the occupation of the SMT level, or
equivalently the partition between the SMT electron and hole.

The spectral function has been found to have the symmetry as follows,
\begin{eqnarray}
A(\omega,V_{mid},-V_{bias})=A(\omega,V_{mid},V_{bias}),
\end{eqnarray}
and
\begin{eqnarray}
A(\omega=\widetilde{\epsilon}_0-\Delta\omega,V_{mid}=\widetilde{\epsilon}_0/e-\Delta V,V_{bias})\notag\\
=A(\omega=\widetilde{\epsilon}_0+\Delta\omega,V_{mid}=\widetilde{\epsilon}_0/e+\Delta
V,V_{bias}) .\label{A-ef-v}
\end{eqnarray}
Compared to the EPI-free case, the spectral function in the EPI is
generally asymmetric with respect to $\widetilde{\epsilon}_0$
unless $eV_{mid}=\widetilde{\epsilon}_0$. This implies a broken
electron-hole symmetry, which can also be seen from
Fig.\ref{A-mu}, where the spectral function in (a) is symmetric
with respect to $\widetilde{\epsilon}_0$; while in Fig.\ref{A-mu}
(b), (c) and (d), there are broken symmetry for electron and hole,
particularly obvious for cases of (c) and (d).

With increasing temperature, one can expect that the profile of
the dot spectral function should be less and less sensitive to the
variation of the lead chemical potentials, because the Fermi
distribution varies continuously across $\mu_{L(R)}$ at high
temperature, so do the phonon sidebands.

\section{The transport properties}
Based on the spectral function discussed above, in this section,
we shall investigate the differential conductance and tunnelling
current of the SMT(QD) system. The case of zero temperature will
be discussed first, then follows the modifications by the finite
temperature effect.

\subsection{The zero temperature}
At zero temperature, the Fermi distribution functions are the step
functions $\Theta(\mu_{L(R)}-\omega)$, and the coefficients $L_n$
reduce to the Eq.\ref{Ln0}, thus the current can be expressed explicitly
as
\begin{widetext}
\begin{eqnarray}
J=\frac{e}{4h}\sum_{n=0}^{\infty}L_n\Gamma\int d\omega \left[
\Theta\left(\mu_L-\omega \right) -\Theta\left(\mu_R-\omega
\right)\right]\left\{ [\Theta\left(\mu_L-\omega-n\omega_0
\right)+\Theta\left(\mu_R-\omega-n\omega_0
\right)]\widetilde{A}(\omega+n\omega_0)\right.\notag\\
\left.+\left[2- \Theta\left(\mu_L-\omega+n\omega_0 \right)
-\Theta\left(\mu_R-\omega+n\omega_0
\right)\right]\widetilde{A}(\omega-n\omega_0)\right\}.
\label{J-zero}
\end{eqnarray}
\end{widetext}
Using $\mu_{L(R)}=eV_{mid}\pm eV_{bias}/2$, and
$\partial\Theta(\mu_{L(R)}-\omega)/\partial V_{bias}=\pm
e\delta(\mu_{L(R)}-\omega)/2$, the differential conductance can be
obtained by performing $\partial J/\partial V_{bias}$,
\begin{widetext}
\begin{eqnarray}
G(V_{mid},V_{bias})=\frac{e^2}{8h}\sum_{n=0}^{\infty}L_n\Gamma\left\{\Theta(eV_{bias}-n\omega_0)\right.
\left[\widetilde{A}(\mu_L)+\widetilde{A}(\mu_R)+
\widetilde{A}(\mu_L-n\omega_0)+\widetilde{A}(\mu_R+n\omega_0)\right]\notag\\
+\Theta(-eV_{bias}-n\omega_0)
\left[\widetilde{A}(\mu_L)+\widetilde{A}(\mu_R)+\widetilde{A}(\mu_L+n\omega_0)\left.+\widetilde{A}(\mu_R-n\omega_0)\right]\right\}.\label{conductance_zero_temperature}
\end{eqnarray}
\end{widetext}
It is easy to verify that the tunnelling current and differential
conductance satisfy the symmetry relations as
\begin{eqnarray}
J(V_{mid}=\widetilde{\epsilon}_0/e-\Delta V,V_{bias})
=J(V_{mid}=\widetilde{\epsilon}_0/e+\Delta V,V_{bias}),\notag\\
G(V_{mid}=\widetilde{\epsilon}_0/e-\Delta V,V_{bias})
=G(V_{mid}=\widetilde{\epsilon}_0/e+\Delta V,V_{bias}),\notag
\end{eqnarray}
and
\begin{eqnarray}
J(V_{mid},-V_{bias})&=&-J(V_{mid},V_{bias}),\notag\\
G(V_{mid},-V_{bias})&=&G(V_{mid},V_{bias}).
\end{eqnarray}
It is noticed that due to the equivalence between the current of
electron type and hole type, the broken symmetry with respect to
$\widetilde{\epsilon}_0$ in the spectral function of SMT electron
is now restored for the tunnelling current and differential
conductance.

\begin{figure}[tbp]
\includegraphics[width=8.5cm]{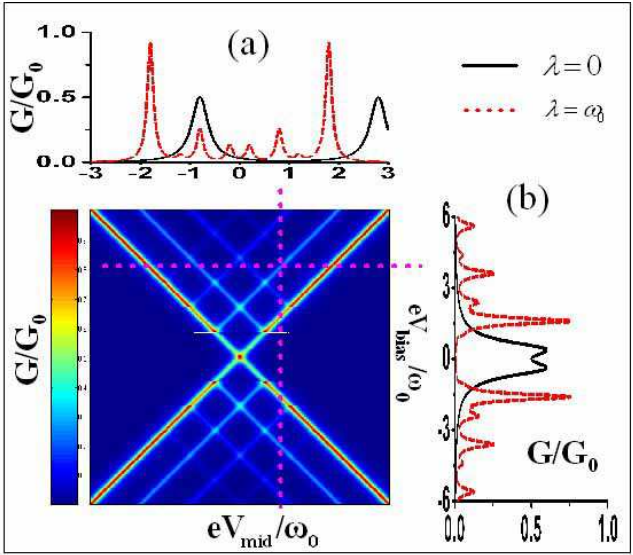}
\caption{(Color on line) The dimensionless differential
conductance ($G_0=e^2/2h$) at $T=0K$ as functions of $V_{mid}$ and
$V_{bias}$. The parameters for calculation are taken as:
$\Gamma=0.2\omega_0$, $\epsilon_0=\omega_0$, and
$\widetilde{\epsilon}_0=0$ which implies that for the sake of
clarity a stronger EPI strength, $\lambda=\omega_0$, is chosen.
The color scale runs from zero (blue) to $G_0$ (red). The above
(a) and right (b) panels to the map are the sections for a fixed
value of $eV_{bias}^{fix}=3.6\omega_0$ and
$eV_{mid}^{fix}=0.8\omega_0$, respectively, marked on the map with
dash lines.} \label{G-ef-vbias}
\end{figure}

\begin{figure}[tbp]
\includegraphics[width=8.5cm]{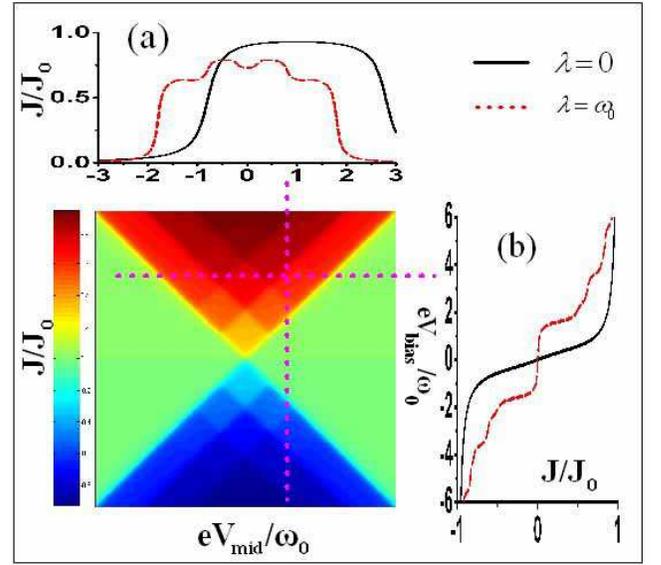}
\caption{(Color on line) The map of tunnelling current,
$J(V_{mid},V_{bias})$, at $T=0K$, where the parameters
 are taken as the same as the Fig.\ref{G-ef-vbias}, and
$J_0=e\Gamma/2\hbar$. The color scale runs from $-0.9J_0$ (blue)
to $0.9J_0$ (red). The solid and dashed lines in sections (a) and
(b) represent the tunnelling current for the cases without and
with EPI, respectively.} \label{J-ef-vbias}
\end{figure}

The map plotted in Fig.\ref{G-ef-vbias} is the calculated
differential conductance as function of $V_{mid}$ and $V_{bias}$
in the presence of the EPI. The dashed lines in
Fig.\ref{G-ef-vbias}(a) and (b) represent the differential
conductance for a fixed value of $eV_{bias}^{fix}=3.6\omega_0$ and
$eV_{mid}^{fix}=0.8\omega_0$, respectively; while the solid lines
correspond to the differential conductance without the EPI as for
reference, in which the peaks appear only when one of the lead
chemical potentials is aligned with the SMT level, ${\it i.e.}$,
$\mu_L(\mu_R)=\epsilon_0$. Compared with the EPI-free case, the
corresponding peaks in the presence of finite EPI are sharpened,
and red-shifted by $g\omega_0$, and  more noticeably, new
satellite peaks are shown up, which are associated with the
phonon-assisted tunnelling processes as depicted in
Fig.\ref{A-mu}(b), when $\mu_{L}=\widetilde{\epsilon}_0+n\omega_0$
or $\mu_{R}=\widetilde{\epsilon}_0-n\omega_0$ for $V_{bias}\geq
0$; $\mu_{L}=\widetilde{\epsilon}_0-n\omega_0$ or
$\mu_{R}=\widetilde{\epsilon}_0+n\omega_0$ for $V_{bias}<0$, where
$n\leq\Theta(|eV_{bias}|/\omega_0)$.

The map plotted in Fig.\ref{J-ef-vbias} is the tunnelling current
as function of $V_{mid}$ and $V_{bias}$, which can be divided into
several plateaus with the bounded lines corresponding to the
differential conductance peaks in Fig.\ref{G-ef-vbias}. It should
be pointed out that the phonon-assisted tunnelling can take place
even at zero temperature. Although no thermal phonon is available
at $0K$, the phonon-emitted process accompanying the tunnel is
possible, if the phonon energy can be supplied by the bias voltage
across the SMT. As also shown in the map (Fig.\ref{G-ef-vbias}),
the phonon-assisted peaks are absent in the left and right
quarters, which are bounded by the lines of the zero-phonon peaks.

Now let us discuss the relationship between the transport
properties and the SMT spectral function. As mentioned above, due
to the dressing effects of EPI, the SMT electron still has a
finite probability to occupy the phonon sidebands even at zero
temperature. Once a phonon sideband of the SMT enters the bias
region, the phonon-assistant-channel would be open, as depicted in
Fig.\ref{A-mu}(b). Thus the information of the SMT spectral
function can be inferred from the spectra of the tunnelling
current or differential conductance. For example, for a fixed
$V_{mid}$, the integral of the differential conductance over
$V_{bias}$ satisfies a sum rule, $\int_{-\infty}^{\infty}dV_{bias}
G(V_{mid},V_{bias})= J_0$, which just results from the
conservation of the spectral weights in Eq.(\ref{sum_A}). While
for a fixed $V_{bias}$, the integral of the differential
conductance over $V_{mid}$ does not equal $J_0$ in consequence of
the transfer of a finite spectral weights to those phonon
sidebands, which do not participate in tunnelling. This is a
little different from the tunnelling process without the EPI,
where the sum rule for the differential conductance always holds.

We will end this subsection with a comparison between the results
based on different decoupling approximations, ${\it i.e.}$,
Eq.(\ref{approximation_G^<}) and Eq.(\ref{approximation_G^r}). In
the zero bias limit, by using Eq.(\ref{approximation_G^<}), the
differential conductance would have no phonon-assisted peaks as
clearly shown in Fig.\ref{G-ef-vbias}. On the other hand, by
following Eq.(\ref{approximation_G^r}), there would be a set of
non-vanishing phonon-assisted peaks in the differential
conductance\cite{Lundin02,Zhu03, Mahan00, Huag96, Meir94}. This
discrepancy can be explained in terms of the different SMT
spectral functions obtained in different approximations. Once the
Green function is decoupled according to
Eq.(\ref{approximation_G^r}), the SMT spectral function would be
expanded in the following way,
\begin{eqnarray}
B(\omega)&=&\sum_{n=-\infty }^{\infty
}L_{n}\left[\widetilde{A}(\omega -n\omega _{0})\right].\label{B}
\end{eqnarray}
At zero temperature, we have $n\ge 0$, the profile of $B(\omega)$ is
similar to Fig.\ref{A-mu}(d), namely there is a set of
non-vanishing phonon sidebands above the zero-phonon band, which
is independent of the Fermi distributions in the leads. Thus,
following the approximation (\ref{approximation_G^r}), in the zero
bias limit, the differential conductance would exhibit a set of
phonon peaks. Physically, our results which follow Eqs.
(\ref{approximation_G^<}) and (\ref{approximation_G^>}),
 seem more reasonable, because the tunnelling electron can neither absorb nor
emit any phonon in the zero temperature and zero bias limit.

\begin{figure}[tbp]
\includegraphics[width=8.5cm]{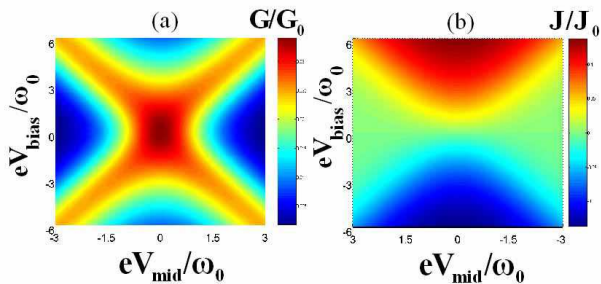}
\caption{(Color on line) The dimensionless differential
conductance and tunnelling current
 as functions of $V_{mid}$ and $V_{bias}$, (a)
$G(V_{mid},V_{bias})$, (b) $J(V_{mid},V_{bias})$, where
$\Gamma=0.2\omega_0$, $\lambda=\omega_0$, $T=0.6\omega_0$,
$\widetilde{\epsilon}_0=0$, and the units $G_0=e^2/2h$,
$J_0=e\Gamma/2\hbar$. The color scale runs from zero conductance
(blue) to $0.05G_0$ (red) in (a) and from $-0.15J_0$ (blue) to
$0.15J_0$ (red) in (b).} \label{JGT}
\end{figure}

\subsection{finite temperature}
The tunnelling current at finite temperature can readily be
obtained from Eq.(\ref{J}).  With the help of  the identities
$L_{-n}=e^{-\beta n\omega_0}L_n$ and $\partial f_{L(R)}(\omega) /
\partial V_{bias}=\pm e\beta
f_{L(R)}(\omega)(1-f_{L(R)}(\omega))/2$, the differential
conductance can be expressed in a compact form:
\begin{eqnarray}
G=\frac{e^2\Gamma}{2h}\sum_{n=-\infty}^{\infty}L_n\int_{-\infty}^{\infty}d\omega
F_n(\omega)\widetilde{A}(\omega-n\omega_0),\label{con-T}
\end{eqnarray}
where the factors $F_n(\omega)$ depend on the Fermi distributions
in the leads through,
\begin{widetext}
\begin{eqnarray}
F_n(\omega)&=&\frac{\beta}{2}\left\{[f_L(\omega)(1-f_L(\omega))+f_R(\omega)(1-f_R(\omega))][1+\frac{e^{-\beta
n\omega_0}-1}{2}(f_L(\omega-n\omega_0)+f_R(\omega-n\omega_0))]\right.\notag\\
&+&\left.\frac{e^{-\beta n\omega_0}-1}{2}(f_L(\omega)-f_R(\omega))
[f_L(\omega-n\omega_0)(1-f_L(\omega-n\omega_0))-f_R(\omega-n\omega_0)(1-f_R(\omega-n\omega_0))]\right\}.\label{Fn}
\end{eqnarray}
\end{widetext}

Both maps for the differential conductance and tunnelling current
at finite temperature are plotted in Fig.\ref{JGT}. Compared to
the zero temperature case, the differential conductance peaks at
finite temperature are broadened and smeared out to some extent,
and the tunnelling current profile become smoother and less
sensitive to the bias, which results from the smoother Fermi
distributions in the leads at finite temperature. Moreover, since
the phonon sidebands of the spectral function will distribute more
symmetrically around the zero-phonon-peak at finite temperature
than the $0K$ case as shown in Fig.\ref{A-lambda-T}, the
phonon-assisted peaks of the differential conductance might appear
in the two forbidden quarters as shown in Fig.\ref{G-ef-vbias} at
$0K$. Therefore, the difference caused by the different decoupling
approximations as discussed above will be diminished with
increasing the temperatures.

\section{conclusion}
In summary, we have theoretically studied the non-equilibrium
quantum transport through the single molecule transistor or
quantum dot in the presence of the local electron-phonon
interaction. Owing to the EPI, the spectral function of the dot
electron will manifest itself in the phonon-dressing effects, such
as the red-shift and sharpening of the zero phonon peak, and the
transfer of finite spectral weights to the phonon sidebands.
Furthermore, due to the interplay between the EPI and the dot-lead
coupling, the zero-phonon band and the phonon sidebands of the
spectral function at low temperature critically depend on the
chemical potentials at two leads and can thus be manipulated.
Although the sum of the spectral weights for the $+n$-th and
$-n$-th phonon sideband is fixed for a given temperature and EPI
strength, the distribution of the spectral weights is dependent on
the chemical potentials in two leads, in particular on the average
of two lead potentials relative to the renormalized SMT level,
showing the broken electron-hole symmetry. The tunnelling current
and differential conductance have been analyzed at both zero and
finite temperatures, which also reveal the dependence on the
chemical potentials of the leads. Different approximations used in
decoupling the electron-phonon system have been compared and
discussed. It has been found that although different decoupling
approximations can yield nearly the same results at high
temperature, they do lead to quite different behaviors at zero
temperature.

{\it{Acknowledgement:}} The authors would like to thank Mr. Hui
Zhai, Mr. Hai-Zhou Lu, Chao-Xing Liu, Ren-bao Liu and Prof. Li
Chang, Tai-Kai Ng for useful discussions and suggestions. This
work is supported by the Natural Science Foundation of China
(Grant No. 90103027, 10374056), the MOE of China (Grant No.200221,
2002003089), and the Program of Basic Research Development of
China (Grant No. 2001CB610508).

\end{document}